\documentclass{article}

\usepackage[utf8]{inputenc}
\usepackage[T1]{fontenc}
\usepackage{lmodern}
\usepackage{natbib}
\usepackage{amsmath}
\usepackage{siunitx}
\usepackage{longtable}
\usepackage{ifthen}
\usepackage[english]{babel}
\usepackage{csquotes}
\usepackage{graphicx} 
\usepackage[font=small,labelfont=bf]{caption} 
\usepackage{url}
\usepackage{subcaption}
\usepackage{authblk}
\usepackage{booktabs} 
\usepackage{amsmath}
\usepackage{amssymb}
\usepackage{hyperref} 
\DeclareMathOperator{\alr}{alr}

\newcommand{\airbnbp}{Airbnb}

\title{Forecasting the U.S. Renewable-Energy Mix with an ALR-BDARMA
Compositional Time-Series Framework}
\author[1]{Harrison Katz}
\author[2]{Thomas Maierhofer}
\affil[1]{Forecasting, Data Science, \airbnbp}
\affil[2]{Department of Statistics \& Data Science, UCLA}
\date{} 

\begin{document}
\maketitle

\begin{abstract}
Accurate forecasts of the U.S. renewable energy consumption mix are essential for planning transmission upgrades, sizing storage, and setting balancing market rules. We introduce a Bayesian Dirichlet ARMA model (BDARMA) tailored to monthly shares of hydro, geothermal, solar, wind, wood, municipal waste, and biofuels from January 2010 through January 2025.
The mean vector is modeled with a parsimonious VAR(2) in additive log ratio space, while the Dirichlet concentration parameter follows an intercept plus five Fourier harmonics, allowing seasonal widening and narrowing of predictive dispersion.
Forecast performance is assessed with a 61 split rolling origin experiment that issues twelve month density forecasts from January 2019 to January 2024.
Compared with three alternatives, a Gaussian VAR(2) fitted in transform space, a seasonal naive that repeats last year's proportions, and a drift free ALR random walk, BDARMA lowers the mean continuous ranked probability score by 15 to 60 percent, achieves component wise 90 percent interval coverage near nominal, and maintains point accuracy (Aitchison RMSE) on par with the Gaussian VAR through eight months and within 0.02 units afterward.
These results highlight BDARMA's ability to deliver sharp, well calibrated probabilistic forecasts for multivariate renewable energy shares without sacrificing point precision. Supplementary Material: An electric power only robustness analysis aligned with the EIA STEO baseline (strict vintaging) is reported in Sections S1 and S2.

\end{abstract}


\section{Introduction}\label{sec:intro}

Electric‑sector decarbonization hinges not only on expanding renewable
output but also on anticipating \emph{how the mix of generation
technologies will evolve}.  Hydropower, wind, solar, biomass and
geothermal differ sharply in marginal cost, intermittency and siting
constraints; reliable medium‑term \emph{mix forecasts} therefore shape
transmission expansion, storage sizing and market design
\citep{IEA2024,EIA2024,EIAfaq2023}. Renewables already supply about
one‑fifth of U.S.\ utility‑scale electricity and their share is expected
to double before 2050, making the coherence and accuracy of share
forecasts more important than ever.

Shares are \emph{compositional}: they are bounded between zero and one
and must sum to unity.  Forecasting each component in isolation, as is
common with univariate ARIMA or machine‑learning regressions
\citep{PanapakidisDagoumas2016,ChenEtAl2017}, yields incoherent
predictions that may turn negative or exceed 100\%
\citep{HyndmanAthanasopoulos2018}. Aitchison’s log‑ratio geometry
provides a principled fix \citep{Aitchison1986}. Early multivariate
illustrations, such as the VAR for geological compositions in
\citet{BillheimerGuttorpFong2001}, and the state‑space model of
\citet{Snyder2017}, demonstrate that standard Gaussian machinery works
once data are mapped to real space.  Still, Gaussian log‑ratio models
often overstate predictive dispersion and ignore seasonally varying
volatility.

Recent research therefore models the composition itself.  A Dirichlet
ARMA process was proposed by \citet{Zheng2017}, a dynamic
Dirichlet–multinomial filter by \citet{KoopmanLeeLucas2023}, and a deep
hierarchical Dirichlet forecaster by \citet{Das2023}.  In the cross‑sectional
domain, \citet{Morais04072018} show that Dirichlet and compositional
regression can outperform traditional attraction models when explaining
brand market shares, underscoring the versatility of simplex‑based
methods.  A direct antecedent to the present study is the Bayesian
Dirichlet ARMA (BDARMA) framework \citep{Katz2024,katz2025bayesiandirichletautoregressiveconditional}, and
subsequently explored with shrinkage priors for trading‑sector shares by
\citet{forecast7030032}, which we adopt here as the data model for a new
application to the U.S.\ renewable‑energy mix.

Applications to energy shares remain limited but growing.  Compositional
VAR and ARMA models, and more recently regional optimization
studies, have been used to project national and sub‑national energy
structures in China, the USA and Canada
\citep{Wei2021,He2022,Xu2024,Xiao2023}. Grey‑system and hybrid
approaches, such as adaptive discrete grey models and MGM–BPNN–ARIMA
designs for broad‑mix or bio‑energy forecasting, further boost accuracy
while respecting the simplex constraint
\citep{Qian2022,Zhang2022,Suo2024}. Machine‑learning work such as the
LSTM study by \citet{Ma2018} and logistic growth analysis of U.S.\ energy
trajectories by \citet{Harris2018} underline the need to tame
nonlinearities, but they still rely on ad‑hoc renormalization.

We apply the Bayesian Dirichlet ARMA framework to the seven‑component
U.S.\ renewable‑energy mix measured monthly from 2010 to 2024, a data set
with pronounced seasonality and secular trends hitherto unaddressed in
the Dirichlet literature.  Forecast skill is benchmarked against three
alternatives: a Gaussian VAR(2) in additive‑log‑ratio space
with identical Fourier dummies, a seasonal naïve that repeats the mix
observed twelve months earlier, and a drift‑free ALR random walk. A
61‑split rolling protocol produces 732 out‑of‑sample density forecasts
and shows that the Dirichlet model attains the strongest probabilistic performance while 
maintaining the VAR’s point accuracy. A full-sample forecast to early 2026 projects 
wind and solar surpassing one-third of renewable generation, providing a coherent picture 
for transmission and storage planning.

The remainder of the paper is organized as follows.
Section~\ref{sec:data} describes the \textsc{EIA} data and seasonal
covariates. Section~\ref{sec:models} presents the BDARMA and benchmark
models. Section~\ref{sec:eval} details the rolling evaluation
protocol and scoring rules. Results are discussed in
Section~\ref{sec:results}, and Section~\ref{sec:conclusion} concludes
with policy implications and avenues for future research. A complementary robustness analysis aligned with the EIA STEO definitions is presented 
in the Supplementary Material (Secs. S1–S2).

\section{Data}\label{sec:data}

The empirical analysis relies on the \textsc{EIA} monthly
\emph{renewable‑energy consumption} data set.  We retain \(T = 181\)
consecutive months from January 2010 through January 2025.  Each
observation is a seven‑part composition
\(
  \mathbf y_t = (y_{t,\text{hyd}}, y_{t,\text{geo}}, y_{t,\text{sol}},
                 y_{t,\text{win}}, y_{t,\text{woo}}, y_{t,\text{was}},
                 y_{t,\text{bio}})^\top
  \in \mathcal S_7,
\)
where shares are obtained by dividing each raw series by their monthly
total.

\paragraph{Additive‑log‑ratio (ALR) coordinates.}
Throughout we analyse the seven‑part composition in \emph{additive‑log‑ratio} form
\begin{align}
    \label{eq:alr_def}  
    \mathbf e_t = \operatorname{alr}(\mathbf y_t)
    = \bigl(\log\tfrac{y_{t,\text{hyd}}}{y_{t,\text{bio}}},
            \log\tfrac{y_{t,\text{geo}}}{y_{t,\text{bio}}},
            \log\tfrac{y_{t,\text{sol}}}{y_{t,\text{bio}}},
            \log\tfrac{y_{t,\text{win}}}{y_{t,\text{bio}}},
            \log\tfrac{y_{t,\text{woo}}}{y_{t,\text{bio}}},
            \log\tfrac{y_{t,\text{was}}}{y_{t,\text{bio}}}\bigr)^\top
  \in\mathbb R^{6},
\end{align}
where \emph{biofuels} serve as the common denominator (reference part).
The inverse map \(\operatorname{alr}^{-1}:\mathbb R^{6}\!\to\!\mathcal S_7\) restores a share vector via
\(
  y_{t,j}= \exp(e_{t,j})\Bigl[1+\sum_{k=1}^{6}\exp(e_{t,k})\Bigr]^{-1}
\)
for \(j\le6\) and
\(y_{t,\text{bio}}=\bigl[1+\sum_{k=1}^{6}\exp(e_{t,k})\bigr]^{-1}\).
We write \(e_{t,j}\) for the \(j\)-th ALR coordinate and
collect them as \(e_1,\dots ,e_6\) when no time index is needed.

\paragraph{Electric‑power–only benchmarking.}
For comparisons to the \textsc{EIA}–\textsc{STEO} industry baseline we also
construct an electric‑power–only view (hydro, geothermal, \emph{solar = utility‑scale
+ small PV}, wind, wood, waste) with monthly closure to the simplex.
Implementation details and the strict vintaging rule are in the
Supplementary Material (Sec.~S1).

\subsection{Exploratory data analysis}\label{subsec:eda}

\begin{figure}[ht]
  \centering
  \includegraphics[width=.82\textwidth]{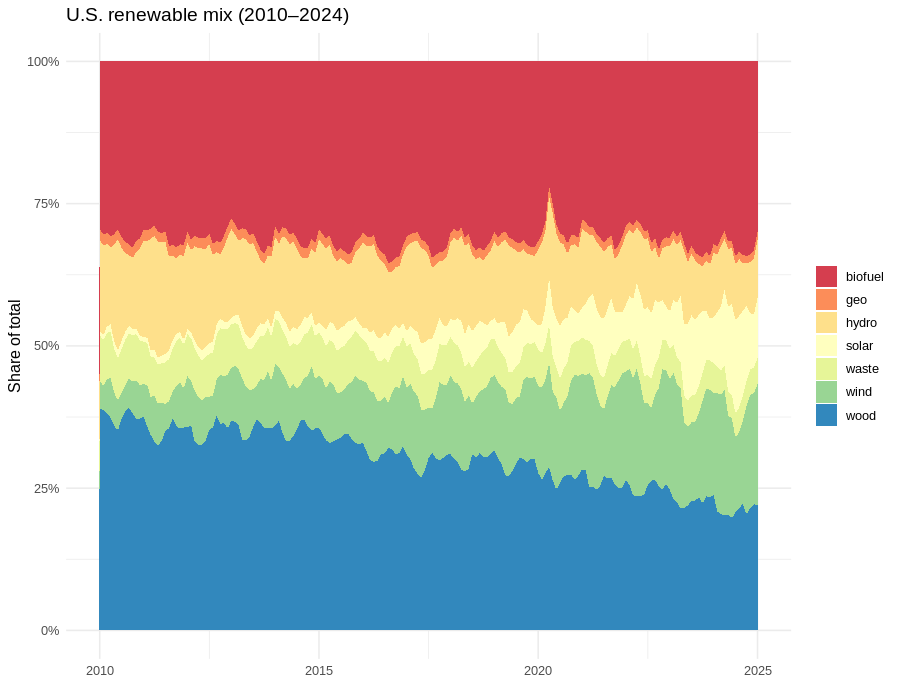}
  \caption{Monthly U.S.\ renewable‑energy mix, 2010–2025.
           Hydro loses ground, while wind and solar expand rapidly.
           Seasonality is most pronounced in hydro (spring runoff) and
           wind (winter–spring peak).  Areas are stacked so each month
           sums to 100\%.}
  \label{fig:history}
\end{figure}

Figure~\ref{fig:history} highlights two macro‑patterns in the sample:  
\textit{(i)} pronounced, asymmetric intra‑annual seasonality and  
\textit{(ii)} a medium‑run reallocation of market share from hydro to
wind and solar.

\begin{figure}[ht]
  \centering
  \subfloat[Seasonal box plots by component\label{fig:eda_box}]{%
    \includegraphics[width=.7\textwidth]{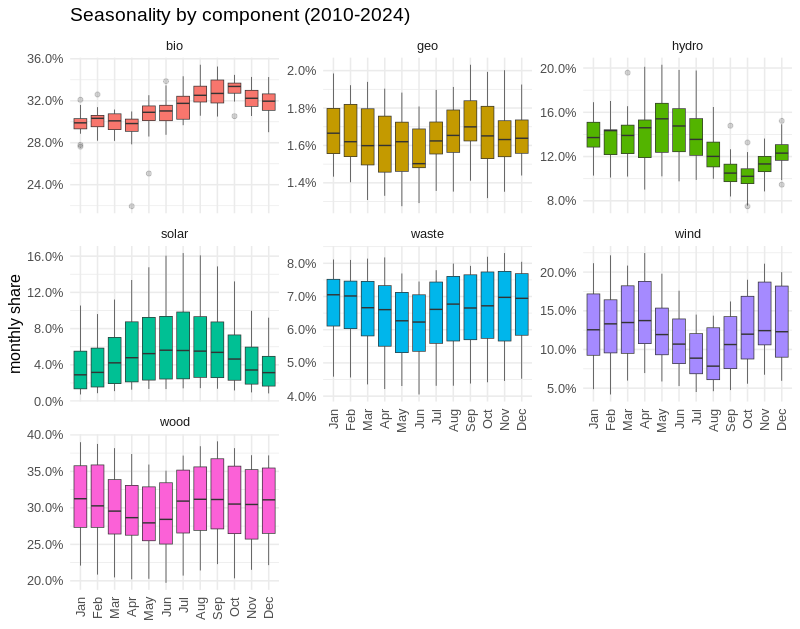}}
  \hfill
  \subfloat[Average intra‑year profile\label{fig:eda_profile}]{%
    \includegraphics[width=.7\textwidth]{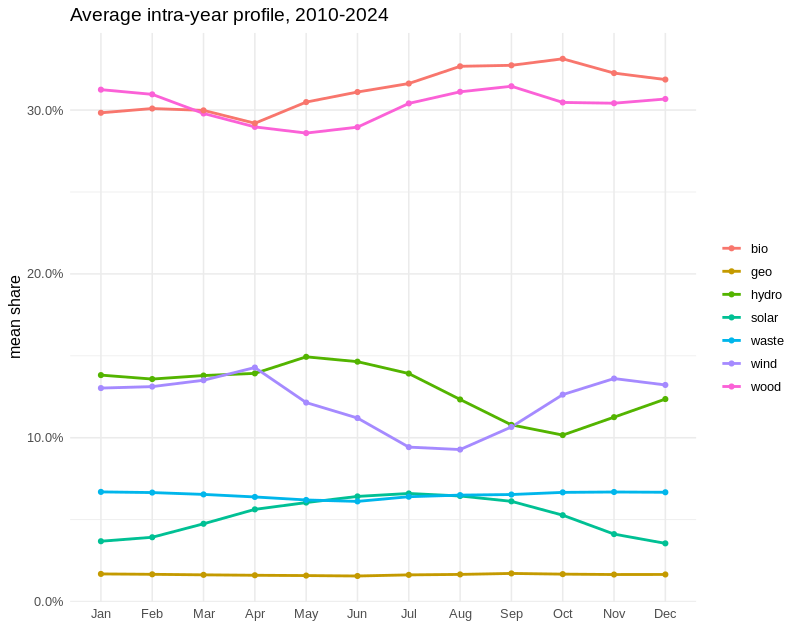}}
  \caption{Seasonal variation in renewable‑energy shares, 2010–2024.
           Boxes span the inter‑quartile range; black bars mark the
           median. Means in panel (b) highlight opposing hydro/solar and
           hydro/wind peaks.}
  \label{fig:eda_season}
\end{figure}

Panel (a) of Figure~\ref{fig:eda_season} shows component‑wise box plots
of monthly shares for 2010–2024; panel (b) traces the mean intra‑year
profile. Hydro exhibits the largest seasonal swing, peaking in
April–May and troughed in late summer while wind follows a bimodal
winter/autumn pattern and solar the mirror image with a July plateau.
Biomass, geothermal and waste are comparatively flat, with median
intra‑year movements below 1 pp. 

\begin{figure}[ht]
  \centering
  \subfloat[Correlation matrix of ALR coordinates\label{fig:eda_corr}]{%
    \includegraphics[width=.7\textwidth]{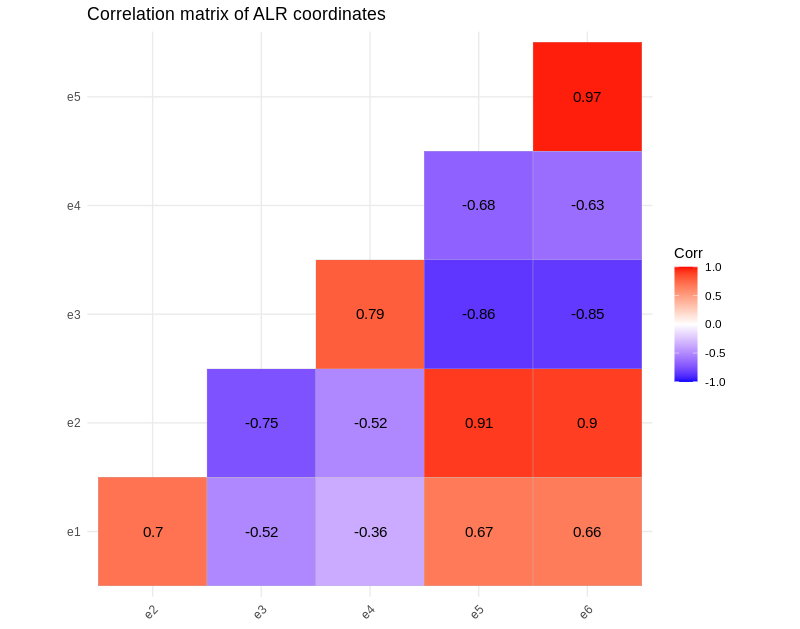}}
  \hfill
  \subfloat[Pairwise ALR scatter plots\label{fig:eda_pairs}]{%
    \includegraphics[width=.7\textwidth]{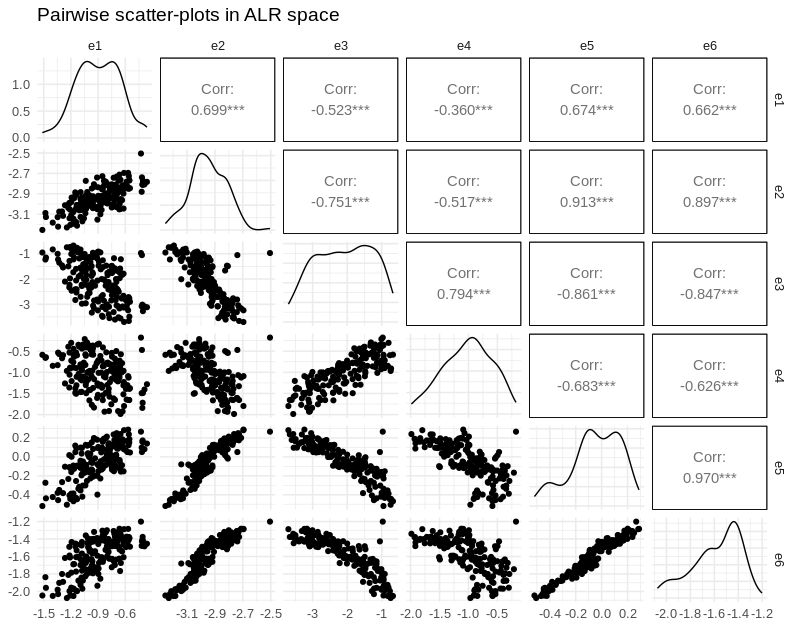}}
  \caption{Cross‑source dependence in additive‑log‑ratio space.  Positive
           (red) and negative (blue) correlations in panel (a) exceed
           0.9 in absolute magnitude; scatter plots in panel (b) reflect
           the non‑linear shape induced by the simplex geometry.}
  \label{fig:eda_correlation}
\end{figure}

Figure~\ref{fig:eda_correlation}\,(a) plots the correlation matrix of
the six ALR coordinates $e_1,\dots,e_6$ (biofuels as reference).  Solar and wind move almost one‑for‑one relative to biofuels
($\rho_{e_3,e_4}=0.97$), whereas hydro and wind are strongly
anti‑correlated ($\rho_{e_1,e_4}\approx-0.86$). Panel (b) confirms these
pair‑wise relations are non‑linear, displaying the characteristic
banana‑shaped clouds induced by log‑ratio geometry.


Table~\ref{tab:sumstats} shows wide dispersion differences: hydro
ranges from \(7.5\%\) to \(20.3\%\) (SD \(=2.6\)\,pp), geothermal is
quasi‑deterministic (SD \(=0.17\)\,pp) and wood the most volatile
component (SD \(=5.1\)\,pp).  

\begin{figure}[ht]
  \centering
  \includegraphics[width=.9\textwidth]{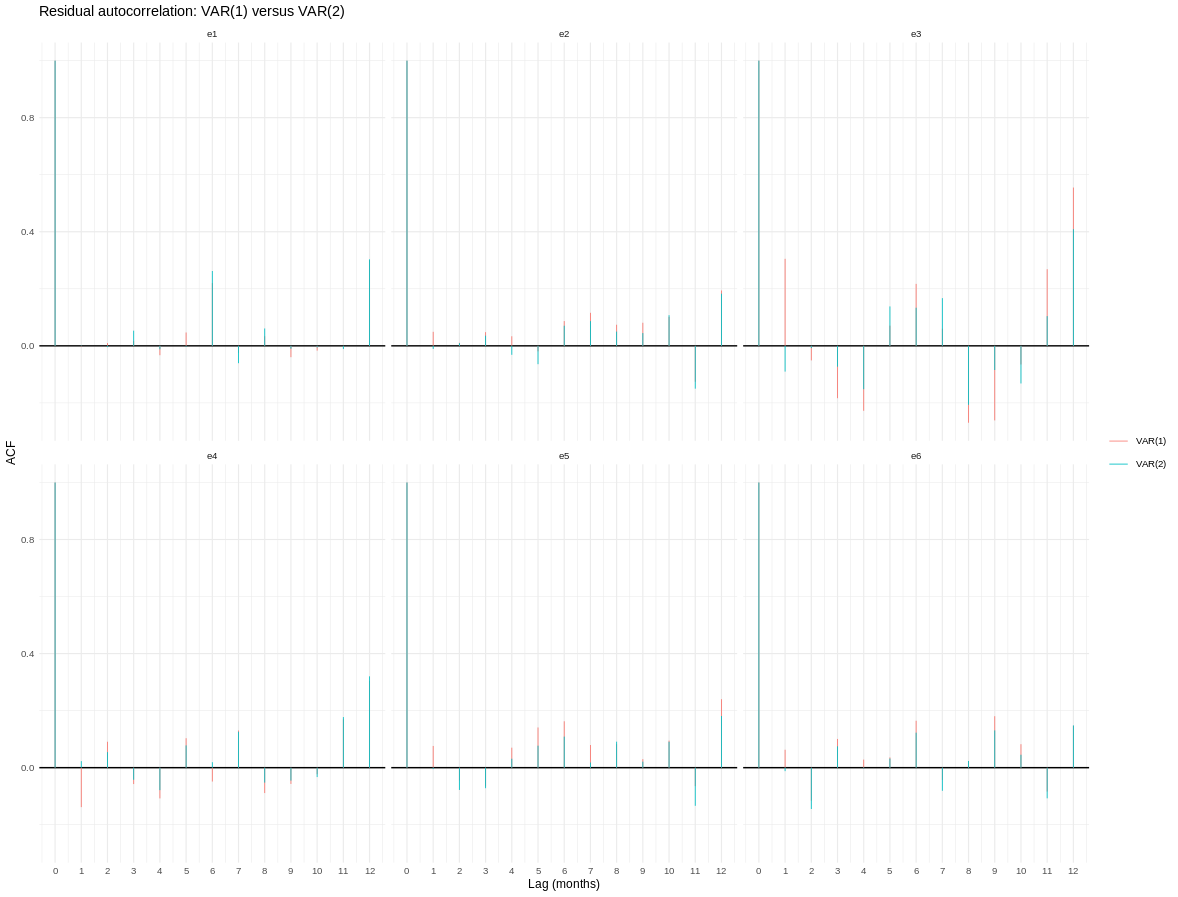}
  \caption{Residual autocorrelation by ALR coordinate:
           red=VAR(1); blue=VAR(2). Adding the second lag removes
           the large spikes at lags 1–2.}
  \label{fig:acf_resid}
\end{figure}

To determine the minimum dynamic order in ALR space we fitted
VAR(1) and VAR(2) models and applied Ljung–Box
and Hosking portmanteau tests to the residuals
(Table~\ref{tab:var_diag}). A Ljung–Box residual diagnostic rejects the white‑noise null for coordinate $e_3$ under VAR(1) (p < 0.001), whereas no coordinate is rejected under VAR(2) (smallest p‑value=.14).  The residual ACF panels in Figure~\ref{fig:acf_resid} show that the prominent spikes at lags 1–2
present under VAR(1) vanish when the second lag is added.  At the
system level, the portmanteau statistic at horizon 12 remains
marginally significant; adding centered monthly dummies reduces
\(\chi^{2}\) from 628 to 431 (\(p=0.006\)). Because VAR(2) is the
\emph{smallest} specification to clear all short‑run autocorrelation and
further lags inflate the parameter count without material gain, we adopt
a VAR(2) mean and address any residual seasonality through exogenous
Fourier terms.


\begin{table}[ht]
  \centering
  \caption{Component means and dispersion, 2010–2024
           (percent of total renewables).}
  \label{tab:sumstats}
  \small
  \begin{tabular}{lccccccc}
    \toprule
          & Hydro & Geo & Solar & Wind & Wood & Waste & Bio \\ \midrule
    Mean (\%) & 13.0 & 1.64 & 5.20 & 12.2 & 30.3 & 6.51 & 31.2 \\
    SD   (\%) &  2.63 & 0.17 & 3.87 &  4.65 &  5.10 & 1.15 & 1.86 \\ \midrule
    Q1  (\%) & 10.9 & 1.50 & 1.91 &  8.48 & 26.4 & 5.54 & 30.1 \\
    Q3  (\%) & 14.8 & 1.77 & 7.53 & 15.1  & 35.1 & 7.48 & 32.5 \\ \midrule
    Min (\%) &  7.51 & 1.28 & 0.73 &  4.18 & 19.7 & 4.05 & 22.0 \\
    Max (\%) & 20.3 & 2.03 & 16.3 & 22.5  & 39.1 & 8.31 & 35.4 \\ 
    \bottomrule
  \end{tabular}
\end{table}

\begin{table}[ht]
  \centering
  \caption{Residual diagnostic statistics (lags 1–2 for Ljung–Box,
           horizon 12 for Hosking portmanteau).}
  \label{tab:var_diag}
  \small
  \begin{tabular}{lccccccc}
    \toprule
    Model & Test & $e_1$ & $e_2$ & $e_3$ & $e_4$ & $e_5$ & $e_6$ \\ \midrule
    VAR(1) & Ljung–Box $p$ & 0.99 & 0.79 & \textbf{0.00} & 0.08 & 0.49 & 0.20 \\
    VAR(2) & Ljung–Box $p$ & 0.99 & 0.98 & 0.47 & 0.73 & 0.57 & 0.14 \\[2pt]
    \multicolumn{2}{l}{Portmanteau $\chi^{2}$ / $p$} &
      \multicolumn{6}{c}{$628 \; / \;<0.001$ (VAR(2))} \\
    \bottomrule
  \end{tabular}
\end{table}

\begin{figure}[ht]
  \centering
  \includegraphics[width=.9\textwidth]{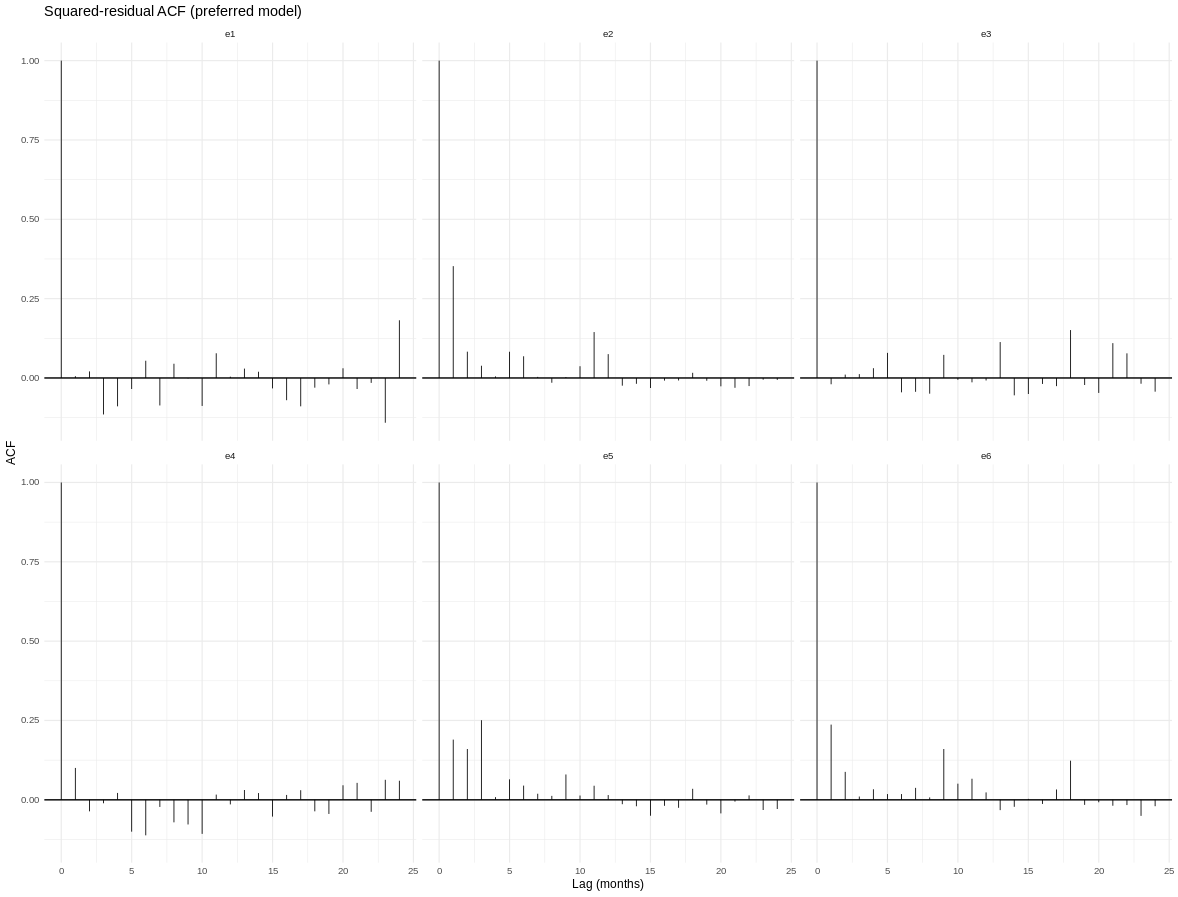}
  \caption{ACF of squared residuals for the preferred
           VAR(2)\,+\,season model. }
  \label{fig:sq_acf}
\end{figure}

\bigskip
\noindent
Because the data exhibit markedly different seasonal amplitudes, strong yet uneven cross‑correlations, and heterogeneous marginal variability, and because residual diagnostics indicate that two lags are the minimum needed for whiteness, we adopt a specification with three complementary elements: a second‑order vector autoregressive mean in ALR space to capture short‑run dynamics; a single seasonal precision curve, common to all components, that modulates forecast dispersion across the calendar year; and a Dirichlet observation model that enforces the compositional sum‑to‑one constraint. Under this Dirichlet layer with a common precision scalar $\phi_t$, component‑wise variances differ only through their mean shares—$\mathrm{Var}(y_{t,j}\mid \mu_t,\phi_t)=\mu_{t,j}\bigl(1-\mu_{t,j}\bigr)/(\phi_t+1)$—rather than via component‑specific precision processes.

All computations were carried out in \textsf{R}~4.3.2 with Stan~2.33 via the \texttt{cmdstanr} interface.
\citep{Stan2024rstan,StanManual2023}.  
Data wrangling, graphics and tables relied on
\texttt{tidyverse} \citep{Wickham2019Tidyverse},
\texttt{lubridate} \citep{Grolemund2011Lubridate},
\texttt{janitor} \citep{Firke2023Janitor},
\texttt{scales} \citep{Wickham2019Scales},
\texttt{patchwork} \citep{Pedersen2025Patchwork},
\texttt{ggcorrplot} \citep{Kassambara2022Ggcorrplot},
\texttt{GGally} \citep{Schloerke2021GGally} and
\texttt{kableExtra} \citep{Zhu2021KableExtra}.
Compositional methods used
\texttt{compositions} \citep{Boogaart2024Compositions} and the
\texttt{transport} package for Aitchison norms
\citep{Schuhmacher2020Transport}.  
Time‑series estimation and testing employed
\texttt{vars} \citep{Pfaff2008Vars},
\texttt{FinTS} \citep{Pfaff2024FinTS} and
\texttt{MTS} \citep{Tsay2023MTS}.

\section{Forecasting model}\label{sec:models}

Let the monthly renewable‑energy mix be the \(J\!=\!7\)-component composition  
\[\mathbf y_t=\bigl(y_{t,\text{hyd}},y_{t,\text{geo}},y_{t,\text{sol}},y_{t,\text{win}},
                    y_{t,\text{woo}},y_{t,\text{was}},y_{t,\text{bio}}\bigr)^{\!\top}\in\mathcal S_7,\;
 t=1,\dots,T.\] Biofuels (\(j^{\!*}=7\)) serve as reference part in every additive‑log‑ratio (ALR) transform that follows.

We model \(\mathbf y_t\) as a Dirichlet distributed whose parameter
vector factorizes into a simplex‑valued mean \(\boldsymbol\mu_t\) and a positive precision scalar \(\varphi_t\):
\begin{equation}
y_t \mid \mu_t, \phi_t \sim \mathrm{Dirichlet}\!\big(\phi_t \mu_t\big), 
\quad \mu_t \in \mathcal{S}_7,\;\phi_t>0.
\label{eq:dirichlet}
\end{equation}

Let \(\boldsymbol\eta_t=\operatorname{alr}(\boldsymbol\mu_t)\in\mathbb R^{J-1}\);
for \(J=7\) this is a six‑vector
\(\boldsymbol\eta_t=(\eta_{t1},\dots,\eta_{t6})^{\!\top}\) of log‑ratios
against biofuels.  Its inverse is
\[
\mu_{tj}=
  \frac{\exp(\eta_{tj})}{1+\sum_{k=1}^{6}\exp(\eta_{tk})}\;(j\le6),
  \qquad
\mu_{t,j^{\!*}}=
  \Bigl[1+\textstyle\sum_{k=1}^{6}\exp(\eta_{tk})\Bigr]^{-1}.
\]

Calendar variation in forecast dispersion is captured by letting the
log‑precision depend on an intercept and five Fourier harmonics (ten sine/cosine terms):
\begin{equation}
  \log\phi_t
    =\mathbf f_t^{\!\top}\boldsymbol\gamma,
  \quad
  \mathbf f_t=(1,\mathbf g_t^{\!\top})^{\!\top},\;
  \mathbf g_t=\bigl(\sin\tfrac{2\pi t}{12},\cos\tfrac{2\pi t}{12},
          \ldots,
          \sin\tfrac{10\pi t}{12},\cos\tfrac{10\pi t}{12}\bigr)^{\!\top},
  \quad
  \boldsymbol\gamma\in\mathbb R^{11}.
\label{eq:precision}
\end{equation}

Short‑run cross‑technology interactions are modelled with a
second‑order vector autoregression process in ALR space:
\begin{equation}
  \boldsymbol\eta_t
  =\mathbf X_t\boldsymbol\beta
   +\mathbf A_1\bigl(\boldsymbol\eta_{t-1}-\mathbf X_{t-1}\boldsymbol\beta\bigr)
   +\mathbf A_2\bigl(\boldsymbol\eta_{t-2}-\mathbf X_{t-2}\boldsymbol\beta\bigr),
  \qquad
  \mathbf X_t = I_{J-1}\otimes\mathbf f_t^{\!\top},
\label{eq:state}
\end{equation}
where (i) \(\mathbf A_1,\mathbf A_2\in\mathbb R^{6\times6}\) are AR
coefficient matrices; (ii) \(\mathbf X_t\) block‑replicates the \(11\)-vector
\(\mathbf f_t\) across the six ALR coordinates, giving
\(\mathbf X_t\in\mathbb R^{6\times66}\); and (iii)
\(\boldsymbol\beta\in\mathbb R^{66}\) contains component‑specific regression
slopes for the seasonal dummies.

With a scalar precision $\phi_t$, the Dirichlet implies a restricted covariance:
$\mathrm{Cov}(y_{i},y_{j})=-\mu_{i}\mu_{j}/(\phi_t+1)$ for $i\neq j$, i.e., negative off‑diagonals of fixed shape. Cross‑component comovement beyond the unit‑sum constraint therefore enters through the mean dynamics rather than the observation variance. Extensions include generalized Dirichlet or logistic‑normal layers, or component‑specific precisions $\phi_{j,t}$ with regularization; we leave these for future work.

\subsection*{Geometric preliminaries and evaluation mapping}
Let $y_t\in\mathcal{S}_7$ denote the share vector and $e_t=\operatorname{alr}(y_t)\in\mathbb{R}^{6}$
its additive log‑ratio (ALR) coordinates with biofuels as the reference part; $\operatorname{alr}^{-1}$
restores shares (see Eq.~(1)). We \emph{model} the mean in ALR space (Eq.~(4)) and obtain
\emph{predictive draws in share space} from the Dirichlet observation layer (Eq.~\eqref{eq:dirichlet}).
Forecasts are evaluated in two complementary spaces: CRPS in share space for joint sharpness
and calibration (Eq.~(5)), and clr‑based RMSE in Aitchison geometry for point accuracy
(Eq.~(6)). Because space choice induces different cross‑component dependencies, reporting
both clarifies where improvements arise. 

\noindent\textit{Reference‑free coordinates.}
The logistic‑normal (sometimes “ALN”) family places a Gaussian law on log‑ratio
coordinates; the isometric log‑ratio (ILR) transform provides orthonormal, reference‑free
coordinates with full metric equivalence on the simplex \citep{AitchisonShen1980,Egozcue2003,Aitchison1986}. This means that the DARMA data models with these three link functions are equivalent, provided the same transformation is applied to the priors. We retain ALR for interpretability and continuity with Eq.~(1).

Each scalar element of \(\mathbf A_1,\mathbf A_2,\boldsymbol\beta\) and
\(\boldsymbol\gamma\) receives an independent
\(\mathcal N(0,1)\) prior.  Posterior inference proceeds via Hamiltonian
Monte Carlo (four chains; 500 warm‑up and 500 retained iterations per
chain) in \textsf{Stan}, yielding \(2\,000\) draws that underpin all
density‑forecast evaluations presented later.


\subsection{Transform‑space VAR(2) (\textbf{tVAR(2)})}

Working in ALR coordinates,
\[
  \boldsymbol\eta_t=
  \mathbf F_{1}\boldsymbol\eta_{t-1}
 +\mathbf F_{2}\boldsymbol\eta_{t-2}
 +\mathbf X_t\boldsymbol\delta
 +\boldsymbol\varepsilon_t,\qquad
 \boldsymbol\varepsilon_t\sim\mathcal N(\mathbf 0,\boldsymbol\Sigma).
\]
Parameters \((\mathbf F_1,\mathbf F_2,\boldsymbol\delta,\boldsymbol\Sigma)\)
are estimated by ordinary least squares with the same seasonal
regressors \(\mathbf X_t\).  Multi‑step forecasts are generated under
the Gaussian innovation assumption and mapped back with
\(\alr^{-1}\).

\subsection{Additive‑log‑ratio random walk (\textbf{ALR–RW})}

A drift‑free benchmark sets each future ALR vector equal to the most
recent observation:
\(
  \boldsymbol\eta_{t+h\mid t}=\boldsymbol\eta_{t}.
\)
Back‑transformation yields a single point forecast with zero predictive
spread.

\subsection{Seasonal naïve copy‑last‑year (\textbf{S‑NAIVE})}

The seasonal naïve copies the composition observed 12 months earlier:
\(
  \mathbf y_{t+h\mid t}=\mathbf y_{t+h-12}.
\)

These four specifications exploit the same information set but differ in
how they propagate seasonality, cross‑technology dependence and
uncertainty. Section~\ref{sec:eval} details the rolling protocol
used to compare their point and density‑forecast performance.

\section{Forecast–evaluation protocol}\label{sec:eval}

Model comparison follows an expanding–window, rolling‑origin design that
mirrors the workflow used by system operators and energy planners. Let $\tau_{s}$, $s=1,\dots,S$, denote the final observation included in
estimation window $s$ and let $H$ denote the fixed forecast horizon
($H=12$).  The first origin is $\tau_{1}=\text{2019‑01}$ and the last
origin that still admits a twelve‑step look‑ahead is
$\tau_{S}=\text{2024‑01}$, so $S=61$.  At origin $s$ the estimation set
is $\{\mathbf y_t:1\le t\le\tau_{s}\}$ while the verification set
comprises $\{\mathbf y_{\,\tau_{s}+h}:h=1,\dots,H\}$.

\subsection{Generating predictive distributions}\label{subsec:pred_dists}

All four competitors are evaluated on Monte‑Carlo samples of equal size \(M=2{,}000\) so that scoring rules are comparable.

\paragraph{BDARMA.}
For every origin \(s\) we retain the \(M\) posterior draws 
\(
\{\boldsymbol\theta^{(m)}\}_{m=1}^{M}
\)
returned by the Hamiltonian Monte‑Carlo sampler.  
Each draw is propagated through the deterministic state equation
\eqref{eq:state} for \(h=1{:}H\) steps, producing the latent
mean \(\boldsymbol\mu^{(m)}_{s,h}\); a single realization
\(
\mathbf y^{(m)}_{s,h}\sim\text{Dirichlet}(\phi^{(m)}_{s,h}\boldsymbol\mu^{(m)}_{s,h})
\)
is then generated from the observation density \eqref{eq:dirichlet}.
The empirical set
\(
  \mathcal P^{\text{BDARMA}}_{s,h}
  =\{\mathbf y^{(m)}_{s,h}\}_{m=1}^{M}
\)
constitutes the predictive distribution.

\paragraph{tVAR(2).}
Let \(\widehat{\boldsymbol\eta}_{s,h}\) and
\(\widehat{\mathbf V}_{s,h}\) be, respectively, the conditional mean and
covariance of the Gaussian forecast for the ALR vector at horizon
\(h\).  
We draw
\(
\boldsymbol\eta^{(m)}_{s,h}\sim
\mathcal N(\widehat{\boldsymbol\eta}_{s,h},\widehat{\mathbf V}_{s,h})
\),
transform with \(\text{alr}^{-1}\), and obtain
\(
  \mathcal P^{\text{tVAR}}_{s,h}
  =\{\text{alr}^{-1}(\boldsymbol\eta^{(m)}_{s,h})\}_{m=1}^{M}.
\)
Multi‑step forecasts are generated under the Gaussian‑innovation assumption and then mapped back to shares with \(\operatorname{alr}^{-1}\), which preserves unit‑sum coherence.

\paragraph{ALR random walk (ALR–RW).}
The point forecast is the last observed ALR vector
\(\boldsymbol\eta_{s,0}\).  
To give the model a distribution that can be scored with CRPS we set
\(
\boldsymbol\eta^{(m)}_{s,h}=\boldsymbol\eta_{s,0}
\)
for every \(m\) and
define
\(
  \mathcal P^{\text{RW}}_{s,h}
  =\{\text{alr}^{-1}(\boldsymbol\eta_{s,0})\}_{m=1}^{M}.
\)
The resulting cloud is degenerate but has the same cardinality \(M\).

\paragraph{Seasonal naïve (S‑NAIVE).}
For each horizon \(h\) we copy the composition observed exactly one
year earlier, \(\mathbf y_{s-12+h}\).  
As with the random walk we replicate this deterministic vector \(M\)
times,
\(
  \mathcal P^{\text{S–NAIVE}}_{s,h}
  =\{\mathbf y_{s-12+h}\}_{m=1}^{M}.
\)

\subsection{Scoring rules}

Denote by $\mathbf y_{s,h}$ the realized share vector at lead $h$
originating from window $s$. Two proper scoring rules are applied.

\paragraph{Energy score (multivariate CRPS).}
Writing
$\|\mathbf a\|_1=\sum_{j=1}^7|a_j|$ for the $\ell_1$ norm, the
sample‑based energy score (ES; a multivariate generalization of the CRPS) is
\begin{equation}
  \operatorname{ES}_{s,h}(\mathcal P)
  = \frac1{M}\sum_{m=1}^{M}
      \bigl\|\mathbf y_{\,s,h}^{(m)}-\mathbf y_{s,h}\bigr\|_{1}
    -\frac1{2M^{2}}\sum_{m=1}^{M}\sum_{m^{\prime}=1}^{M}
      \bigl\|\mathbf y_{\,s,h}^{(m)}-
              \mathbf y_{\,s,h}^{(m^{\prime})}\bigr\|_{1}.
  \label{eq:crps}
\end{equation}
We use the $\ell_1$ norm so that units are “share points”; ES remains a strictly proper scoring rule under common norms.

\paragraph{Aitchison root‑mean‑square error.}
Let $\widehat{\boldsymbol\mu}_{s,h}=M^{-1}\sum_{m}
\mathbf y_{\,s,h}^{(m)}$ be the posterior mean.
With the centred log‑ratio
$\operatorname{clr}(\mathbf p)=
 (\log p_1/g,\ldots,\log p_7/g)$ and
geometric mean $g=(\prod_{j=1}^7p_j)^{1/7}$, the point‑forecast error is
\begin{equation}
  \operatorname{RMSE}_{s,h}
  = \bigl\|\operatorname{clr}(\mathbf y_{s,h})
          -\operatorname{clr}(\widehat{\boldsymbol\mu}_{s,h})
     \bigr\|_2/\sqrt{7}.
  \label{eq:rmse}
\end{equation}
Both \eqref{eq:crps} and \eqref{eq:rmse} reduce to zero for a perfect
forecast.

\paragraph{Interval diagnostics.}
For BDARMA the 5‑th and 95‑th sample quantiles define a 90 \% credible
interval for each component.  Coverage is tallied over all
$(s,h)$ pairs.

\subsection{External baseline and scoring in electric‑only space}\label{subsec:steo_eval}
A matched evaluation against the vintaged \textsc{EIA}–\textsc{STEO} baseline in the EP‑only frame uses the same rolling origins, horizons, and scoring rules; see Supplementary Sec.~S1 for construction and Sec.~S2 for the horizon‑by‑horizon results (Figs.~S1–S2; Tables~S1–S2).

\subsection{Fixed‑origin projection}

After the rolling study a single fixed‑origin forecast is produced from
the complete estimation window
2010‑01 – 2025‑01 ($\tau^\ast=T$).  Future Fourier regressors
$\mathbf f_{T+h}$ are generated deterministically, so the only source of
uncertainty is the posterior distribution of model parameters and, for
tVAR(2), the Gaussian state noise. 

\section{Results}\label{sec:results}

\subsection{Forecast accuracy across horizons}\label{subsec:rolling_results}

\begin{figure}[ht]
  \centering
  \includegraphics[width=.82\textwidth]{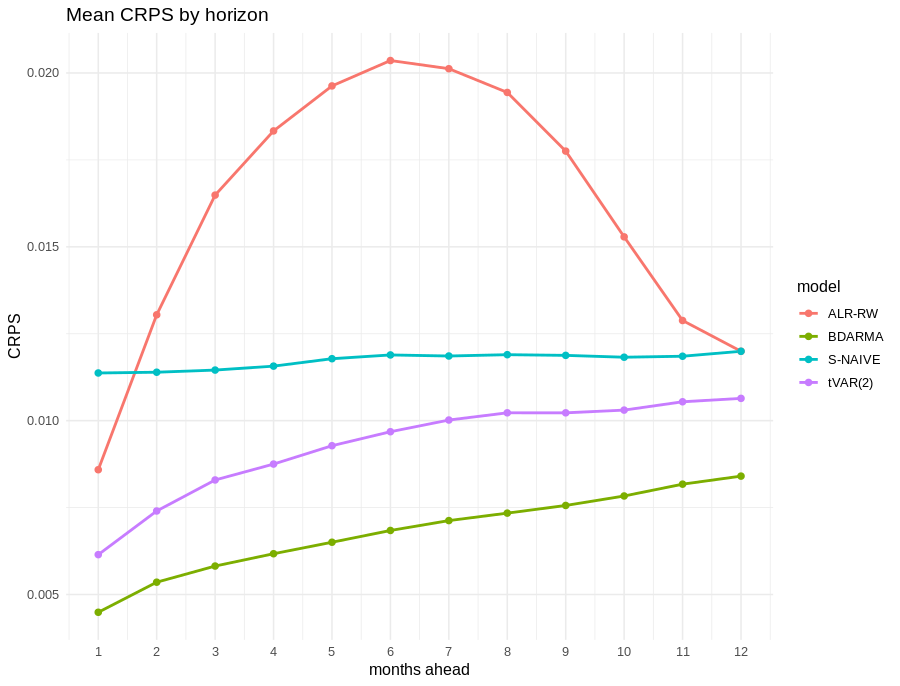}
  \caption{Mean CRPS by horizon, averaged over 61 rolling origins. Lower values indicate sharper and better‑calibrated densities.}
  \label{fig:crps_h}
\end{figure}

\begin{figure}[ht]
  \centering
  \includegraphics[width=.82\textwidth]{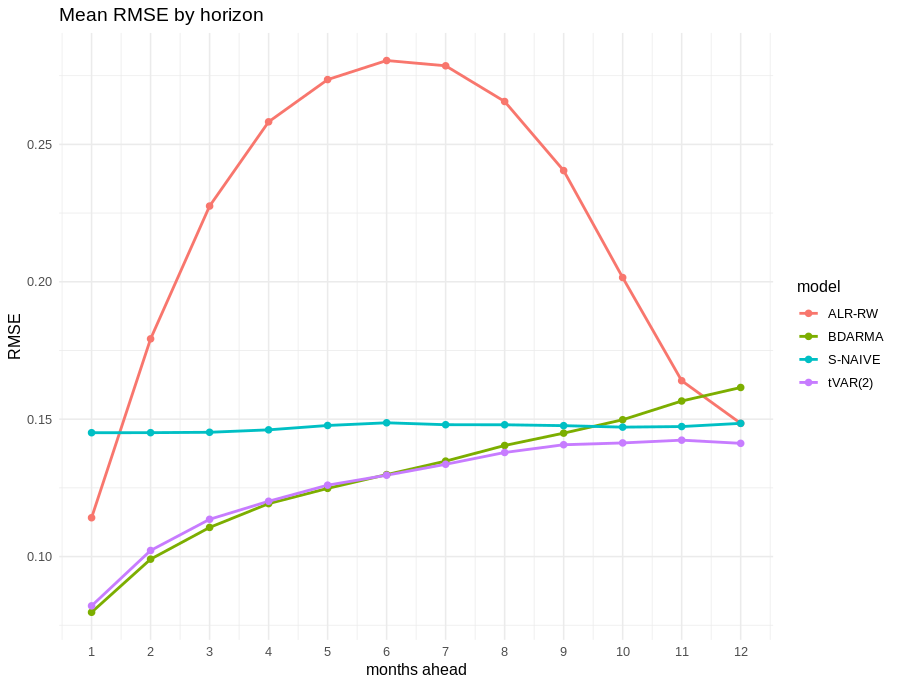}
  \caption{Mean Aitchison RMSE by horizon. Lower values indicate more accurate point forecasts.}
  \label{fig:rmse_h}
\end{figure}

Tables~\ref{tab:crps_h} and~\ref{tab:rmse_h} report mean CRPS and mean Aitchison RMSE by horizon; Figures~\ref{fig:crps_h} and~\ref{fig:rmse_h} visualize the same quantities. Across sixty‑one rolling origins, the Bayesian Dirichlet ARMA (BDARMA) model attains the lowest CRPS at every horizon. At one month it is about one quarter lower than the transform‑space VAR(2) and more than half lower than either naïve rule, and the advantage widens with lead: by twelve months BDARMA still improves on tVAR(2) by roughly one fifth and on S‑NAIVE by about forty percent, while ALR‑RW remains weakest overall. 

Point errors give a complementary view. Through eight months, BDARMA and tVAR(2) yield nearly identical RMSEs; from month nine onward, the Gaussian VAR gains a small edge that peaks at roughly one hundredth of an Aitchison unit. That edge comes with broader predictive spreads, which shows up as persistently higher CRPS for the VAR.

\paragraph{Technology‑specific interpretation.}
Component patterns matter for planning. Wind and hydro exhibit the largest seasonal swings, so their predictive bands are wider and more seasonally structured, informing spring runoff scheduling for hydro and winter ramping reserves for wind. Solar’s long‑run rise with a summer plateau produces medium‑horizon gains that help quantify midday surplus risk and storage sizing. Geothermal and waste behave almost deterministically, supporting narrow tolerance bands for compliance or procurement. Wood remains comparatively volatile across horizons, arguing for conservative hedging where biomass supply or policy constraints bind. These qualitative statements align with Table~\ref{tab:cov_comp}, where geothermal and waste approach perfect inclusion while wind and biofuels are harder to capture due to stronger seasonality and policy or demand variability.

\begin{table}[ht]
\centering
\caption{Mean CRPS across rolling origins. Boldface marks the best score for each horizon.}
\label{tab:crps_h}
\small
\begin{tabular}{rcccc}
\toprule
Horizon & BDARMA & tVAR(2) & S‑NAIVE & ALR‑RW\\
\midrule
1  & \textbf{0.00449} & 0.00615 & 0.0114 & 0.0086\\
2  & \textbf{0.00535} & 0.00740 & 0.0114 & 0.0130\\
3  & \textbf{0.00582} & 0.00829 & 0.0115 & 0.0165\\
4  & \textbf{0.00617} & 0.00875 & 0.0116 & 0.0183\\
5  & \textbf{0.00650} & 0.00928 & 0.0118 & 0.0196\\
6  & \textbf{0.00684} & 0.00968 & 0.0119 & 0.0204\\
7  & \textbf{0.00713} & 0.0100  & 0.0119 & 0.0201\\
8  & \textbf{0.00734} & 0.0102  & 0.0119 & 0.0194\\
9  & \textbf{0.00756} & 0.0102  & 0.0119 & 0.0178\\
10 & \textbf{0.00783} & 0.0103  & 0.0118 & 0.0153\\
11 & \textbf{0.00817} & 0.0105  & 0.0119 & 0.0129\\
12 & \textbf{0.00841} & 0.0106  & 0.0120 & 0.0120\\
\bottomrule
\end{tabular}
\end{table}

\begin{table}[ht]
\centering
\caption{Mean Aitchison RMSE across rolling origins. Boldface marks the lowest error at each horizon.}
\label{tab:rmse_h}
\small
\begin{tabular}{rcccc}
\toprule
Horizon & BDARMA & tVAR(2) & S‑NAIVE & ALR‑RW\\
\midrule
1  & \textbf{0.0797} & 0.0821 & 0.145 & 0.114\\
2  & \textbf{0.0990} & 0.102  & 0.145 & 0.179\\
3  & \textbf{0.111}  & 0.114  & 0.145 & 0.228\\
4  & \textbf{0.119}  & 0.120  & 0.146 & 0.258\\
5  & \textbf{0.125}  & 0.126  & 0.148 & 0.274\\
6  & 0.130           & \textbf{0.130} & 0.149 & 0.281\\
7  & 0.135           & \textbf{0.134} & 0.148 & 0.279\\
8  & 0.140           & \textbf{0.138} & 0.148 & 0.266\\
9  & 0.145           & \textbf{0.141} & 0.148 & 0.240\\
10 & 0.150           & \textbf{0.141} & 0.147 & 0.202\\
11 & 0.157           & \textbf{0.142} & 0.147 & 0.164\\
12 & 0.162           & \textbf{0.141} & 0.148 & 0.148\\
\bottomrule
\end{tabular}
\end{table}

\subsection{Coverage of BDARMA predictive intervals}

The Monte Carlo intervals are well calibrated. Componentwise 90\% coverage rises from 86\% at one month to 99\% by a full year (Table~\ref{tab:cov_hor}). By technology (Table~\ref{tab:cov_comp}), geothermal and waste approach perfect inclusion; solar and hydro are very close to nominal; wind and biofuels are lower, consistent with larger seasonal amplitude and policy‑ or demand‑driven swings.

\begin{table}[ht]
\centering
\caption{Empirical 90 percent coverage of BDARMA component intervals.}
\label{tab:cov_hor}
\small
\begin{tabular}{ccccccccccccc}
\toprule
$h$ & 1 & 2 & 3 & 4 & 5 & 6 & 7 & 8 & 9 & 10 & 11 & 12\\
\midrule
Cov & .863 & .891 & .907 & .933 & .950 & .957 & .963 & .975 & .984 & .980 & .983 & .985\\
\bottomrule
\end{tabular}
\end{table}

\begin{table}[ht]
\centering
\caption{Componentwise BDARMA coverage (61 by 12 forecasts).}
\label{tab:cov_comp}
\small
\begin{tabular}{lccccccc}
\toprule
Hydro & Geo & Solar & Wind & Wood & Waste & Bio\\ \midrule
.939 & 1.000 & .993 & .886 & .954 & .998 & .862\\ \bottomrule
\end{tabular}
\end{table}

\subsection{Fixed‑origin comparison}\label{subsec:fixed_origin_fan}

\begin{figure}[ht]
  \centering
  \includegraphics[scale=.32]{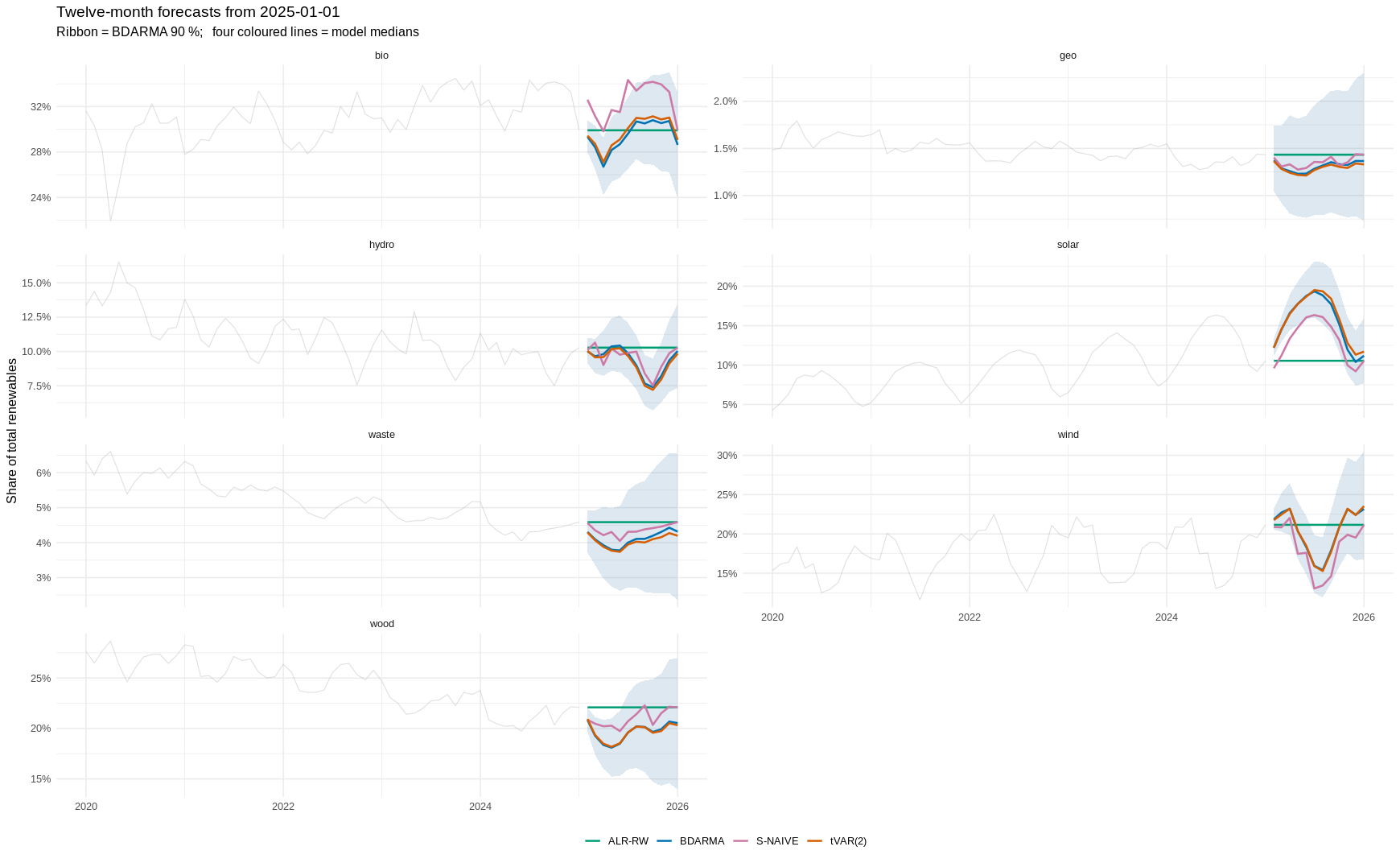}
  \caption{Twelve‑month forecasts issued 2025-01-01 after refitting all models to the 2010–2024 sample. Blue shading denotes the BDARMA 90\% predictive interval. Colored lines are posterior or plug‑in medians. Axes are free by facet.}
  \label{fig:fan_fixed}
\end{figure}

The one‑year trajectories in Figure~\ref{fig:fan_fixed} reinforce these patterns. BDARMA produces calibrated bands that widen where seasonality and trend are strong and tighten where series are stable, supporting a single forecast set for both point planning and risk assessment. The naïve rules supply medians only. The Gaussian VAR yields medians close to BDARMA but cannot indicate whether any remaining gap is material relative to forecast dispersion.

If decisions depend almost entirely on point forecasts, exogenous regressors are abundant and trusted, and speed is critical, the transform‑space VAR(2) is reasonable, especially beyond nine months where it has a slight RMSE edge. The seasonal naïve can suffice for steady components and as a monitoring benchmark where deviations from a baseline are the main signal. BDARMA remains preferable when calibrated densities, compositional coherence, and seasonal uncertainty are central to the decision.

Sequence models such as LSTMs or GRUs can be competitive with long histories and rich covariates. In monthly, medium‑length settings with few components, they require careful constraints to preserve the unit sum and substantial tuning to obtain calibrated densities. Relative to such sequence learners and to gradient‑boosted ensembles, BDARMA offers three practical advantages for this task: (i) coherence by construction (no post‑hoc renormalization), (ii) direct density forecasts in share space rather than ad‑hoc bands, and (iii) strong performance with limited history. The approaches are complementary: neural or boosted summaries of weather or policy can enter BDARMA as exogenous features, and neural forecasts can be used as external signals.

Estimation uses four chains with 1000 warm‑up and draw counts; origins and chains parallelize across cores. Design matrices grow linearly with the number of components and harmonics.

A matched electric‑power‑only comparison against the vintaged \textsc{EIA}–\textsc{STEO} baseline confirms the main findings: BDARMA leads at one month, while \textsc{STEO} is strongest beyond two months. Details are in the Supplementary Material (Sections~S1–S2; Figures~S1–S2; Tables~S1–S2).

\section{Conclusion}\label{sec:conclusion}

This paper develops and evaluates a Bayesian Dirichlet ARMA (BDARMA) model for monthly U.S.\ renewable‑energy shares, with a VAR(2) mean in additive‑log‑ratio space and a seasonal Dirichlet precision. In a 61‑split rolling evaluation, BDARMA delivers the sharpest, best‑calibrated twelve‑month density forecasts: mean CRPS is lower than a transform‑space VAR(2) at every horizon and far below naïve rules, while point accuracy (Aitchison RMSE) matches the VAR through eight months and stays within roughly two hundredths thereafter. The result is improved uncertainty quantification without sacrificing the central path.

The errors have planning implications by technology. Hydro and wind, our most seasonal components, retain wider, seasonally patterned bands that inform spring runoff scheduling and winter ramping reserves. Solar’s trend plus summer plateau yields medium‑horizon gains that tighten estimates of midday surplus risk and storage needs. Geothermal and waste are near‑deterministic, justifying narrow tolerance bands, while wood’s broader dispersion argues for conservative procurement where fuel and policy are more volatile.

A balanced view also clarifies when simpler models are reasonable. VAR(2) can be preferred when decisions hinge almost entirely on point forecasts, exogenous regressors are abundant, and computational speed is paramount; the seasonal naïve is defensible for steady systems and as a monitoring baseline. By contrast, BDARMA should be the default when calibrated densities, coherence across components, and explicit seasonality matter for reserves, transmission, and storage planning.

\section*{Acknowledgments}
Sean Wilson for insightful discussions and for his invaluable assistance in developing the original \emph{BDARMA} Stan code.

\subsection*{Code Availability}
All \texttt{R} scripts and \texttt{Stan} model files used in this study are publicly available at \\
\href{https://github.com/harrisonekatz/energy-compositions}{https://github.com/harrisonekatz/energy-compositions}. 
All results and figures in this manuscript can be reproduced by running the scripts found in that repository.

\subsection*{Conflict of Interest}
The authors declare no conflicts of interest and that all work and opinions are their own and that the work is not sponsored or endorsed by Airbnb

\bibliographystyle{chicago}
\bibliography{references}
\section{Supplementary: Electric-Power-Only Robustness
         against an Industry Baseline (EIA--STEO)}
\label{sec:sup_ep_only}

\noindent\textbf{Purpose and setup.}
We test robustness against an industry baseline under matched definitions and strict vintaging.
All analysis is in an electric-power-only (EP) frame with six components, scored on shares, and free of look-ahead.

\subsection*{Data and construction}

Monthly six-part composition
\[
\mathbf y_t=\bigl(y_{t,\mathrm{hyd}},\,y_{t,\mathrm{geo}},\,y_{t,\mathrm{sol}},\,y_{t,\mathrm{win}},\,y_{t,\mathrm{woo}},\,y_{t,\mathrm{was}}\bigr)^\top
\in\mathcal S_6,
\]
closed each month to sum to one. \emph{Solar} means \emph{utility-scale solar plus small-scale PV}.

Inputs are extracted from monthly \textsc{EIA} \textsc{STEO} base workbooks and consolidated into two CSVs:
\begin{itemize}
  \item \texttt{steo\_ep\_hist\_and\_forecast\_wide\_allvintages.csv}: by vintage and month, utility-scale generation from Table 7d(1) (hydro, geothermal, solar, wind, wood, waste) and \emph{small PV} from Table 7a; includes a historical/forecast flag.
  \item \texttt{forecasts\_only\_categories\_wide\_allvintages.csv}: forecasts-only for the same variables, by vintage.
\end{itemize}

For month \(t\), select the latest vintage that still labels \(t\) as historical. Form EP levels
\[
\begin{aligned}
(\mathrm{hyd},\mathrm{geo},\mathrm{sol},\mathrm{win},\mathrm{woo},\mathrm{was})_t
&=(\mathrm{hyd\_ep},\,\mathrm{geo\_ep},\,\mathrm{sol\_ep}+\mathrm{solar\_pv\_small},\,\mathrm{win\_ep},\,\mathrm{woo\_ep},\,\mathrm{was\_ep})_t,
\end{aligned}
\]
then apply within-month closure to obtain a six-part share vector. Truth uses only historical columns from the STEO workbooks.

For each forecasting origin \(\tau\), choose the latest STEO vintage released on or before \(\tau\).
From that vintage take forecast \emph{levels} for the six components (solar as above), then re-scale to shares by closure.

Truth is built for 2023-01 through 2025-12.
Rolling monthly origins run from 2024-01 forward, subject to leaving a six-month test window.
Evaluation uses only horizons for which the selected vintage provides a forecast; there is no backfilling.

\subsection*{Methods}

Forecasts are produced in log-ratio space and evaluated on the simplex.
Let \(\mathcal C(\cdot)\) denote closure.
We use additive-log-ratio (ALR) coordinates with \(\mathrm{waste}\) as reference:
\[
\mathbf e_t=\operatorname{alr}(\mathbf y_t)=\Bigl(
\log\tfrac{y_{t,\mathrm{hyd}}}{y_{t,\mathrm{was}}},
\log\tfrac{y_{t,\mathrm{geo}}}{y_{t,\mathrm{was}}},
\log\tfrac{y_{t,\mathrm{sol}}}{y_{t,\mathrm{was}}},
\log\tfrac{y_{t,\mathrm{win}}}{y_{t,\mathrm{was}}},
\log\tfrac{y_{t,\mathrm{woo}}}{y_{t,\mathrm{was}}}
\Bigr)^\top\in\mathbb R^{5}.
\]
Forecasts are mapped back via \(\operatorname{alr}^{-1}\) and then closed \(\mathcal C\).

For origin \(\tau\), the training set is \(\{1,\dots,\tau\}\) (truth shares) and the test set is \(\{\tau+1,\dots,\tau+H\}\) with \(H=6\).
Origins with zero overlap between the test window and the chosen vintage's forecast months are skipped.

\paragraph{Models.}
\begin{enumerate}
  \item BDARMA(1,0) in ALR space with seasonal regressors.
        Observation: Dirichlet on \(\mathcal S_6\) with mean \(\mu_t\) and precision \(\phi_t\), yielding simplex-valued predictive draws.
        The mean is linked through ALR with one autoregressive lag and a Fourier seasonal basis (\(K=3\) harmonics); \(\log\phi_t\) uses the same basis.
        Estimation uses Hamiltonian Monte Carlo in \textsf{Stan}/\textsf{cmdstanr}.
  \item \textbf{VAR(1)+Fourier in ALR space.}
        A VAR(\(p=1\)) for \(\mathbf e_t\) with the same \(K=3\) Fourier regressors as exogenous terms captures short-run cross-technology dynamics and seasonality; forecasts are mapped back to shares.
  \item \textbf{ALR-RW.}
        Random-walk baseline in ALR space: the forecast equals the last observed ALR vector (mapped back and closed).
  \item \textbf{Seasonal naive (S-NAIVE).}
        The \(h\)-step-ahead share vector equals the share vector observed \(12-h\) months before the origin (with closure).
  \item \textbf{EIA STEO (strict vintage).}
        Deterministic point baseline from the selected vintage's forecast levels (solar as above), re-closed to the simplex.
\end{enumerate}

Seasonality enters via \(K=3\) harmonics \(\{\sin(2\pi k t/12),\,\cos(2\pi k t/12)\}_{k=1}^{3}\) plus an intercept.
The basis is replicated across ALR coordinates and used identically in the BDARMA mean, BDARMA precision, and VAR exogenous design.

Accuracy is measured in Aitchison geometry:
\begin{itemize}
  \item \textbf{CLR-RMSE} (points): root mean squared distance between \(\operatorname{clr}(\hat{\mathbf y})\) and \(\operatorname{clr}(\mathbf y)\).
  \item \textbf{CLR-CRPS} (densities): predictive draws are mapped to shares, transformed by centered log-ratios (CLR), and univariate CRPS is averaged across coordinates.
\end{itemize}

\begin{itemize}
  \item \emph{Horizon:} \(H=6\). \emph{Origins:} monthly from 2024-01 through 2025-06, restricted to months with strict-vintage STEO coverage.
  \item \emph{Fourier:} \(K=3\) harmonics.
  \item \emph{BDARMA MCMC:} 4 chains; 750 warm-up and 750 retained iterations per chain; \texttt{adapt\_delta}=0.84; \texttt{max\_treedepth}=11.
  \item \emph{Vintaging:} latest STEO vintage \(\le\) origin; no fallback to later releases; no backfilling of missing horizons.
  \item \emph{Guards:} negative or non-finite components are set to zero before closure; all outputs are re-closed to the simplex.
\end{itemize}

\section{Supplementary Results: Electric-Power-Only Robustness
         against the EIA--STEO Baseline}
\label{sec:sup_ep_only_results}

\begin{figure}[ht]
  \centering
  \includegraphics[scale=.6]{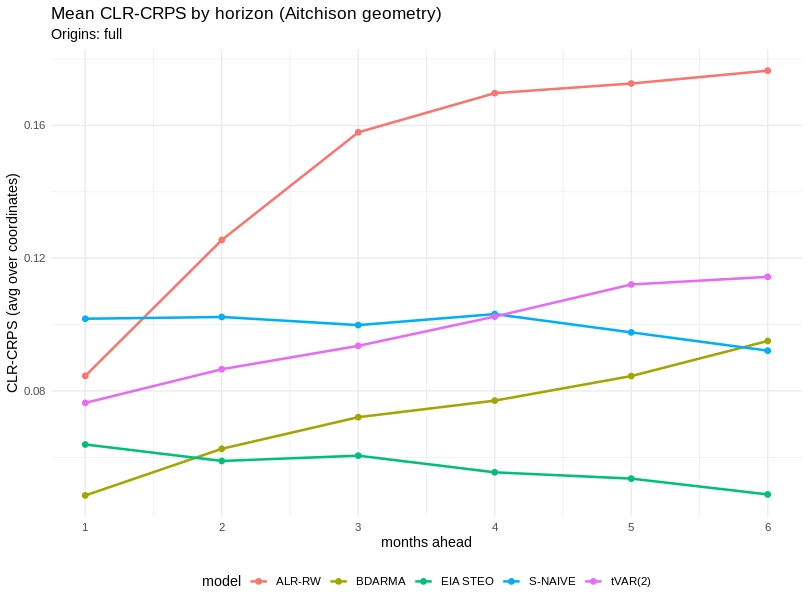}
  \caption{\textbf{Electric-power-only robustness: mean CLR-CRPS by horizon.}
  Average CLR-CRPS (lower is better) for \(h=1,\dots,6\) in a six-technology EP frame
  (hydro, geothermal, solar equals utility-scale plus small PV, wind, wood, waste).
  Rolling monthly origins are 2024-01 to 2025-06; training begins in 2023-01.
  BDARMA is best at \(h=1\); for \(h\ge2\) the strict-vintage \textsc{EIA}--\textsc{STEO} baseline yields the lowest CRPS.}
  \label{fig:ep_crps}
\end{figure}

\begin{figure}[ht]
  \centering
  \includegraphics[scale=.6]{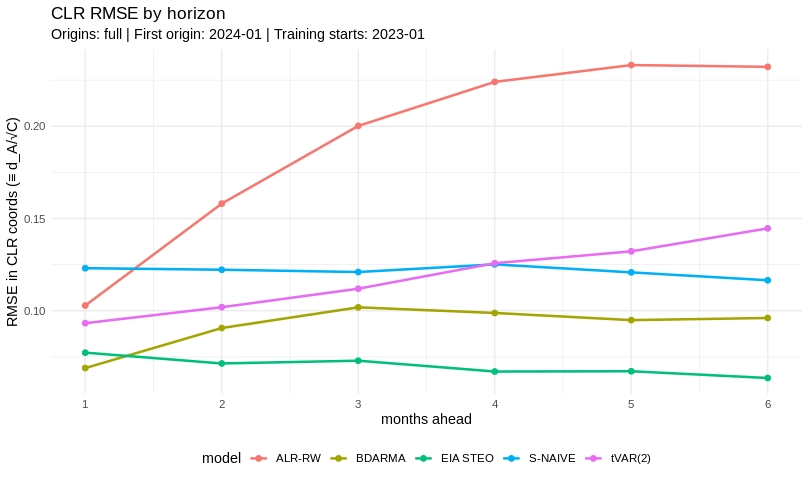}
  \caption{\textbf{Electric-power-only robustness: CLR-RMSE by horizon.}
  Mean CLR-RMSE (lower is better) over the same rolling-origin design as Figure~\ref{fig:ep_crps}.
  Forecasts are produced in ALR space and evaluated after mapping back to shares in Aitchison geometry.
  BDARMA is best at \(h=1\); from \(h=2\) onward the strict-vintage \textsc{EIA}--\textsc{STEO} baseline is smallest.}
  \label{fig:ep_rmse}
\end{figure}

Across monthly rolling origins from 2024-01 through 2025-06 and horizons \(h=1,\dots,6\), Figures~S\ref{fig:ep_crps} and S\ref{fig:ep_rmse} (profiles) are summarized in Tables~\ref{tab:ep_crps} and \ref{tab:ep_rmse}.
At \(h=1\), BDARMA attains the smallest errors (CLR-CRPS \(=0.0485\); CLR-RMSE \(=0.0690\)), improving on the strict-vintage STEO point baseline by about \(24\%\) in CLR-CRPS and \(11\%\) in CLR-RMSE.
From \(h=2\) onward, STEO delivers the lowest errors in both scoring rules and the advantage widens with horizon (for example, at \(h=6\) the STEO CLR-CRPS of \(0.0488\) is about \(49\%\) lower than BDARMA's \(0.0951\), and its CLR-RMSE of \(0.0636\) is about \(34\%\) lower than BDARMA's \(0.0961\)).
The seasonal naive and ALR-RW baselines deteriorate steadily with horizon.
The tVAR(1)+Fourier specification sits between the naive baselines and BDARMA at short horizons and does not overtake STEO at any horizon.

\begin{table}[t]
  \centering
  \caption{Mean CLR-CRPS by horizon (lower is better). Origins: monthly 2024-01 to 2025-06; \(n=12\) origin-months per cell. Best per horizon in \textbf{bold}.}
  \label{tab:ep_crps}
  \small
  \begin{tabular}{ccccccc}
    \toprule
    Horizon & ALR-RW & BDARMA & EIA STEO & S-NAIVE & tVAR(1) \\
    \midrule
    1 & 0.0845 & \textbf{0.0485} & 0.0639 & 0.1020 & 0.0764 \\
    2 & 0.1250 & 0.0626 & \textbf{0.0589} & 0.1020 & 0.0865 \\
    3 & 0.1580 & 0.0721 & \textbf{0.0605} & 0.0998 & 0.0936 \\
    4 & 0.1700 & 0.0771 & \textbf{0.0555} & 0.1030 & 0.1020 \\
    5 & 0.1730 & 0.0845 & \textbf{0.0536} & 0.0976 & 0.1120 \\
    6 & 0.1770 & 0.0951 & \textbf{0.0488} & 0.0921 & 0.1140 \\
    \bottomrule
  \end{tabular}
\end{table}

\begin{table}[t]
  \centering
  \caption{Mean CLR-RMSE by horizon (lower is better). Origins: monthly 2024-01 to 2025-06; \(n=12\) origin-months per cell. Best per horizon in \textbf{bold}.}
  \label{tab:ep_rmse}
  \small
  \begin{tabular}{ccccccc}
    \toprule
    Horizon & ALR-RW & BDARMA & EIA STEO & S-NAIVE & tVAR(1) \\
    \midrule
    1 & 0.1030 & \textbf{0.0690} & 0.0773 & 0.1230 & 0.0933 \\
    2 & 0.1580 & 0.0907 & \textbf{0.0715} & 0.1220 & 0.1020 \\
    3 & 0.2000 & 0.1020 & \textbf{0.0730} & 0.1210 & 0.1120 \\
    4 & 0.2240 & 0.0988 & \textbf{0.0671} & 0.1250 & 0.1260 \\
    5 & 0.2330 & 0.0949 & \textbf{0.0672} & 0.1210 & 0.1320 \\
    6 & 0.2320 & 0.0961 & \textbf{0.0636} & 0.1170 & 0.1450 \\
    \bottomrule
  \end{tabular}
\end{table}

Under strict vintaging and matched definitions, the comparison reflects genuine forecasting differences rather than measurement choices.
The STEO baseline's error profile declines gently with horizon, suggesting a stable medium-term signal in EP shares as defined here.
BDARMA's one-step edge is consistent with dynamic shrinkage toward recent ALR history, and its relative performance tapers with horizon.
The naive alternatives are useful reference points but are dominated across horizons, while tVAR(1)+Fourier remains between BDARMA and the naive baselines.

\end{document}